\documentclass[epj]{svjour}
\usepackage{graphics}
\newcommand{\be}{\begin{equation}}
\newcommand{\bea}{\begin{eqnarray}}
\newcommand{\ee}{\end{equation}}
\newcommand{\eea}{\end{eqnarray}}
\def\chic#1{{\scriptscriptstyle #1}}

\begin{document}
\title{The effective neutrino charge radius}

\author{J. Papavassiliou \and J. Bernab\'eu  \and D. Binosi \and J. Vidal 
}                     

\institute{ Departamento de F\'{\i}sica Te\'orica and IFIC, 
Universidad de Valencia-CSIC, 
E-46100, Burjassot, Valencia, Spain }
\date{2 October, 2003} 
\abstract{
It is shown that at one-loop  order a neutrino charge radius (NCR) may
be  defined, which  is ultraviolet finite, 
does  not depend  on the  gauge-fixing
parameter, nor on properties of the target other than its electric charge.
This  is accomplished through
the systematic decomposition of  physical amplitudes into  effective
self-energies,   vertices,  and   boxes,   which  separately   respect
electroweak gauge invariance.   In this way the NCR  stems solely from
an effective proper photon-neutrino one-loop vertex, which satisfies a
naive, QED-like Ward  identity.  The NCR so defined may
be  extracted  from  experiment,  at  least  in  principle, 
by expressing  a set  of  experimental electron-neutrino
cross-sections in  terms of the  finite NCR and two  additional gauge-
and  renormalization-group-invariant quantities, corresponding  to the
electroweak effective charge and mixing angle.
} 

\maketitle
\section{\label{intro} Introduction}

It is a well-known fact that in non-Abelian gauge theories 
off-shell Green's functions depend explicitly on the 
gauge-fixing parameter. Therefore, the definition of 
quantities familiar from QED, such as effective charges and 
form-factors, is in general problematic.
Such has been the case with the neutrino electromagnetic 
form-factor and the corresponding NCR. 
The calculational fact 
that, within the Standard Model, the (off-shell) 
one-loop $\gamma^{*}\nu \nu$ vertex (and the NCR obtained 
from it) is 
a gauge-dependent quantity has been established beyond any doubt  
in the seventies \cite{Bardeen:1972vi}. 
Based on this observation, it was concluded  
that ``the NCR'', which is the derivative at $q^2 =0$ of the 
electromagnetic form-factor $F(q^2)$ extracted form this vertex,  
is not a physical quantity. Of course, 
if something is gauge-dependent it is not physical. But the fact that 
the off-shell vertex is gauge-dependent only means that it just does not 
serve as a reasonable definition of the NCR, it does not mean that 
an {\it effective} NCR cannot be encountered which satisfies {\it all} necessary  
physical properties, gauge-independence being one of them.  
Indeed, since then,   
several papers in the literature have attempted to find a 
{\it modified} vertex-like amplitude, leading 
to a consistent definition of the electromagnetic 
NCR (see \cite{Bernabeu:2000hf} for an extended list of references). 
The common underlying idea in all such papers is 
to rearrange the Feynman graphs contributing to the 
scattering amplitude 
of neutrinos with charged particles, 
in an attempt to find a vertex-like   
combination that would satisfy all desirable properties. 
Of course, in doing so, a plethora of 
non-trivial physical constraints 
need to be satisfied. For example, one should  
not enforce gauge-independence at the expense of 
introducing target-dependence.  
Therefore, a definite guiding-principle is needed, allowing 
for the construction of physical sub-amplitudes with definite kinematic 
structure (i.e. self-ener- gies, vertices, boxes).
\vspace{-0.4cm}
\section{The pinch technique effective vertex}
\label{gief}
What has been accomplished recently in \cite{Bernabeu:2002nw}
(and some of the literature cited therein) 
is the proof that there exists a 
well-defined and finite effective three-point (vertex) Green's function, 
which has the following properties:
({\bf i}) it is independent of the 
gauge-fixing parameter ($\xi$);
({\bf ii}) it is ultra-violet finite;
({\bf iii})  it satisfies a QED-like Ward-identity;
({\bf iv}) it captures all that is coupled to a genuine $(1/q^2)$ photon propagator;
({\bf v}) it couples electromagnetically to the target;
({\bf vi}) it does not depend on the $SU(2)\times U(1)$ quantum numbers
of the target-particles used;
({\bf vii}) it has a non-trivial dependence on the mass $m_i$ of the 
charged isospin partner $f_i$ of the neutrino in question; 
({\bf viii}) it contains only physical thresholds;
({\bf ix}) it satisfies unitarity and analiticity;
({\bf x}) it can be extracted from experiments.

The theoretical methodology allowing this 
physically meaningful definition is that of the pinch technique (PT) 
\cite{Cornwall:1982zr}. The  PT  is  a
diagrammatic  method which exploits  the  underlying  symmetries 
encoded  in  a  {\it  physical} amplitude  such  as  an  $S$-matrix  element,
in  order  to  construct effective   Green's   functions    with   special  
properties.    
In the context of the NCR, 
the 
basic observation, already put forth in  \cite{Bernabeu:2000hf}, 
is that the gauge-dependent parts of the conventional 
$\gamma^{*}(q)\nu \nu$,  (to which the gauge-dependent $\rm{NCR}$ is associated)
communicate and eventually cancel {\it algebraically}
against analogous  
contributions concealed inside the $Z^{*}(q)\nu \nu$ vertex, the self-energy graphs,
and the box-diagrams (if there are boxes in the process), 
{\it before}   any integration over the 
virtual momenta is carried out. For example, due to rearrangement produced by the 
systematic triggering of elementary Ward identities 
the gauge-dependent contributions 
coming from boxes 
are not box-like, but propagator or vertex-like. 
To understand how the topological modifications, which allow the communication between 
initially different graphs,  
come about, notice that, at one-loop level,  
{\it all}  virtual longitudinal momenta ($k$) originating from tree-level gauge-boson propagators
inside Feynman graphs 
trigger two elementary Ward identities, which furnish {\it inverse propagators}.
The first reads
\begin{eqnarray}
\not\! k  P_{L} &=& (\not\! k + \not\! p ) P_{L} 
- P_{R} \not\! p 
\nonumber\\
&=& S_{f'}^{-1}(\not\! k + \not\! p ) P_{L} - 
P_{ R} S_{f}^{-1}(\not\! p) \nonumber\\
&+& m_{f'} P_{L}  -  m_{f} P_{R},
\label{EWI1}
\end{eqnarray}
where $P_{R(L)} = [1  + (-) \gamma_5]/2$  is the  chirality projection
operator and   
$S_{ f}$ is the tree-level propagator of the fermion $f$;
$f'$ is the
isodoublet-partner of the external fermion $f$.
The second relevant Ward identity reads 
\be
(k+q)^{\nu}\Gamma_{\alpha\mu\nu}(q,k,-k-q) = t_{\alpha\mu}(q) - t_{\alpha\mu}(k) ,  
\label{EWI2}
\ee
where $\Gamma_{\alpha\mu\nu}$ is the bare triple-gauge-boson vertex, and 
$t_{\mu\nu}(q)= q^2 g_{\mu\nu} - q_{\mu}q_{\nu}$. 
We emphasize that {\it all} gauge-dependent parts cancel exactly at the end of the 
pinching procedure, even in the presence of 
{\it non-vanishing fermion masses}
$m_{f}$ and $m_{f'}$, contrary to recent claims \cite{Fujikawa:2003tz}. 

The new one-loop proper three-point function 
$\widehat{\Gamma}^{\mu}_{A \nu_i \bar{\nu}_i}$ 
satisfies the properties listed before. In particular, properties from (iv) to (vi)
ensure that it is a photon vertex, uniquely defined in the sense that it is independent of using
either weak isoscalar sources (coupled to the $B$-field) or weak isovector sources
(coupled to $W^0$), or any charged combination.
The NCR, to be denoted by $\big <r^2_{\nu_i}\,  \big>$, is  
then defined as   
$\big <r^2_{\nu_i}\,  \big> = 6 (d\widehat{F}_{\nu_i}/dq^2)_{q^2=0} $; a straightforward 
calculation yields 
\be
\big <r^2_{\nu_i}\,  \big> =\, 
\frac{G_{\chic F}}{4\, {\sqrt 2 }\, \pi^2} 
\Bigg[3 
- 2\log \Bigg(\frac{m_{\chic i}^2}{M_{\chic W}^2} \Bigg) \Bigg]\, ,
\label{ncr}
\ee
where $i= e,\mu,\tau$, the  
$m_i$ denotes the mass of the charged iso-doublet
partner of the neutrino under consideration, and $G_{\chic F}$
is the Fermi constant. 
\vspace{-0.4cm}
\section{Measuring the effective NCR}
\label{meas}
After arriving at a physically meaningful definition for the 
NCR, the next crucial question is whether the NCR so defined 
constitutes a genuine physical observable. In the rest of this 
section we will briefly discuss 
the method proposed in  \cite{Bernabeu:2002nw}
for the extraction of the NCR from experiment. 

It is important to emphasize that  
measuring the entire process  
$f^{\pm}\nu \to f^{\pm} \nu$ does 
{\it not} constitute a measurement
of the NCR, because by 
 changing the target fermions $f^{\pm}$ one will 
generally change the answer, thus introducing a target-dependence
into a quantity which (supposedly) constitutes an
intrinsic property of the neutrino. 
Instead, what we want to measure is the target-independent
Standard Model NCR only, 
stripped of any contributions depending on the specific properties 
of the target (mass, spin, weak hypercharge), except its electric charge.
Specifically, as mentioned above, the PT
rearrangement of the $S$-matrix
makes possible the definition of distinct, physically meaningful
sub-amplitudes, one of which,   
$\widehat{\Gamma}^{\mu}_{A \nu_i \bar{\nu_i}}$, 
is finite and directly
related to the NCR. 
However, the remaining sub-amplitudes, such as 
self-energy, vertex- and box-corrections,  
even though they 
do not enter into the definition of the
NCR, still contribute numerically to the entire $S$-matrix; 
in fact, some of them combine to form additional  
physical observables of interest, 
most notably the effective (running) 
electroweak charge and mixing angle.
Thus, in order to isolate the NCR, one must conceive of
a combination of experiments and kinematical conditions, 
such that 
all contributions not related to the NCR
will be eliminated.

Consider the elastic processes 
$ f(k_1) \nu(p_1)  \to f(k_2) \nu(p_2) $ and 
$f(k_1) \bar{\nu}(p_1)  \to f(k_2) \bar{\nu}(p_2) $,
where $f$ denotes an electrically charged 
fermion belonging to a different
iso-doublet than the neutrino $\nu$, in order to eliminate 
the diagrams mediated by a charged $W$-boson.
The Mandelstam variables are defined as
$s=(k_1+p_1)^2 = (k_2+p_2)^2$, 
$t= q^2 = (p_1-p_2)^2 = (k_1-k_2)^2$, 
$u = (k_1-p_2)^2 = (k_2-p_1)^2$, 
and $s+t+u=0$. 
In what follows we will restrict ourselves to the 
limit $t=q^2 \to 0$ of the above amplitudes,
assuming that all external (on-shell) fermions are massless.
As a result of this special kinematic situation we have the
following relations:
$p_1^2 = p_2^2 = k_1^2 = k_2^2 = p_1 \cdot p_2 = k_1 \cdot k_2 = 0$
and 
$p_1 \cdot k_1  = p_1 \cdot k_2 = p_2 \cdot k_1 = p_2 \cdot k_2 = s/2 $.
In the center-of-mass system we have that 
$t=-2 E_{\nu}E_{\nu}'(1-x)\leq 0 $, 
where $E_{\nu}$ and $E_{\nu}'$
are the energies of the neutrino before and after the
scattering, respectively, and 
$x \equiv \cos\theta_{cm}$, where
$\theta_{cm}$
is the scattering angle. Clearly, the condition $t=0$ 
corresponds to the exactly forward amplitude, 
with $\theta_{cm}=0$, \, $x=1$.

At tree-level the amplitude  $ f \nu  \to f \nu $ is 
mediated by an off-shell $Z$-boson, coupled to the fermions  
by means of the bare vertex 
$\Gamma_{Z {f} \bar{f}}^{\mu} = -i 
(g_w/ c_w)\, \gamma^{\mu}\, [ v_f + a_f \gamma_5]$
with 
$v_f = s^2_w Q_{f} - \frac{1}{2} T^f_z$  and 
$a_f=\frac{1}{2} T^f_z$.

At one-loop, the relevant  
contributions are determined  
through the PT 
rearrangement of the amplitude, giving rise to 
gauge-independent sub-amplitudes. 
In particular, the one-loop 
$AZ$ self-energy $\widehat{\Sigma}_{\chic{A}\chic{Z}}^{\mu\nu}(q^2)$
obtained is transverse, for {\it both} 
the fermionic and the bosonic contributions,
i.e. $\widehat{\Sigma}_{\chic{A}\chic{Z}}^{\mu\nu}(q^2)
= (q^2 \, g^{\mu\nu}  - q^{\mu} q^{\nu}) 
{\widehat{\Pi}}_{ \chic{A} {\chic Z}} (q^2)$.
Since the external currents are conserved, 
from the $ZZ$ self-energy 
$\widehat{\Sigma}_{\chic{Z}\chic{Z}}^{\mu\nu}(q^2)$
we keep only the part proportional to $g^{\mu\nu}$, 
whose dimension-full cofactor will be denoted 
by $\widehat{\Sigma}_{\chic{Z}\chic{Z}}(q^2)$.
Furthermore, 
the one-loop
vertex 
$\widehat\Gamma_{{\chic Z} {\chic F} \bar{\chic F}}^{\mu}(q,p_1,p_2)$, 
with $F = f$  or $F = \nu$, 
satisfies a QED-like 
Ward identity, relating it to the one-loop  
inverse fermion propagators $\widehat\Sigma_{\chic F}$,
i.e.  
$ q_{\mu} 
\widehat\Gamma_{{\chic Z} {\chic F} \bar{\chic F}}^{\mu}(q,p_1,p_2)
= 
\widehat\Sigma_{\chic F} (p_1) - \widehat\Sigma_{\chic F} (p_2)$.
It is then easy to show that, in the limit of 
$q^2 \to 0$,  
$\widehat\Gamma_{{\chic Z} {\chic F} \bar{\chic F}}^{\mu} 
\sim q^2 \gamma^{\mu}(c_1 + c_2 \gamma_5)$; 
since it is multiplied by a
massive $Z$ boson propagator $(q^2 - M_{\chic Z})^{-1}$, its 
contribution to the amplitude vanishes when 
$q^2 \to 0$. This is to be contrasted with the 
$\widehat{\Gamma}^{\mu}_{A \nu_i \bar{\nu}_i}$,  
which is accompanied by a 
$(1/q^2)$ photon-propagator, thus giving rise 
to a contact interaction between the target-fermion and the neutrino,
described by the NCR. 
 
We next eliminate the box-contributions,
by means of the ``neutrino--anti-neutrino'' method.
The basic observation 
is that the tree-level amplitudes 
${\cal M}_{\nu f}^{(0)}$ 
as well as the
part of the one-loop amplitude ${\cal M}_{\nu f}^{(B)}$
consisting of 
the propagator and vertex corrections 
(namely the ``Born-improved'' amplitude),
are proportional to $[\bar{u}_{f}(k_2)\gamma_{\mu}( v_f + a_f \gamma_5 ) 
u_{f}(k_1)]
[\bar{v}(p_1)\gamma_{\mu} P_{\chic L} \, v(p_2)]$, 
and  
therefore transform differently than the boxes 
under the replacement ~\cite{Sarantakos:1983bp}
$\nu \to \bar{\nu}$, since \footnote{
Eq.(\ref{sar}) appears in Ref.\cite{Bernabeu:2002nw} with 
an inconsequential sign error in the intermediate step.}
\be
\bar{u}(p_2)\gamma_{\mu} P_{\chic L} u(p_1) \to 
- \bar{v}(p_1)\gamma_{\mu} P_{\chic L}  v(p_2)
= - \bar{u}(p_2)\gamma_{\mu} 
P_{\chic R} u(p_1).
\label{sar}
\ee
Thus, under the above transformation, 
${\cal M}_{\nu f}^{(0)} + {\cal M}_{\nu f}^{(B)}$ reverse 
sign once, 
whereas the box contributions reverse sign twice.
These distinct transformation properties 
allow for the isolation of
the box contributions 
when the forward differential cross-sections
$(d\sigma_{\nu f}/dx)_{x=1}$ and  
$(d\sigma_{\bar{\nu} f}/dx)_{x=1}$ 
are appropriately combined. 
In particular, the combination 
$\sigma^{(+)}_{\nu f} \equiv 
(d\sigma_{\nu f}/dx)_{x=1}
+ (d\sigma_{\bar{\nu} f}/dx)_{x=1}$ does not contain 
boxes,
whereas 
the conjugate combination of cross-sections, 
$\sigma^{(-)}_{\nu f} \equiv (d\sigma_{\nu f}/dx)_{x=1}
- (d\sigma_{\bar{\nu} f}/dx)_{x=1}$,   
isolates the contribution of the boxes.

Finally, a detailed analysis shows 
that in the kinematic limit we consider, 
the Bremsstrahlung contribution vanishes, 
due to a completely destructive interference 
between the two relevant  diagrams corresponding to the
processes $f A \nu (\bar{\nu}) \to f \nu (\bar{\nu})$ and 
$f \nu (\bar{\nu}) \to f A \nu (\bar{\nu})$. 
The absence of such corrections is consistent with the
fact that there are no infrared divergent contributions 
from the (vanishing) vertex
$\widehat\Gamma_{{\chic Z} {\chic F} \bar{\chic F}}^{\mu}$, 
to be canceled against.  
 
$\sigma^{(+)}_{\nu f}$ receives contributions from the 
tree-level exchange of a $Z$-boson, the one-loop contributions
from the ultraviolet divergent quantities 
$\widehat{\Sigma}_{\chic{Z}\chic{Z} }(0)$ and  
${\widehat{\Pi}}^{ \chic{A}  {\chic  Z}} (0)$, 
and the (finite) NCR, coming from the proper vertex 
$\widehat{\Gamma}^{\mu}_{A \nu_i \bar{\nu}_i}$.
The first three contributions are universal, i.e. common to all 
neutrino species, whereas that of the proper vertex 
$\widehat{\Gamma}^{\mu}_{A \nu_i \bar{\nu}_i}$
is flavor-dependent. 

To proceed, the renormalization of 
$\widehat{\Sigma}_{\chic{Z}\chic{Z} }(0)$ and  
${\widehat{\Pi}}_{ \chic{A}  {\chic  Z}} (0)$ must be carried out.
It turns out that, by virtue of the Abelian-like Ward-identities 
enforced after the pinch technique rearrangement 
\cite{Cornwall:1982zr},
 the resulting expressions combine in such a way as to form manifestly 
renormalization-group invariant combinations 
\cite{Hagiwara:1994pw}.
In particular, after 
carrying out the standard re-diagonalization,
two such quantities 
may be constructed:
\bea
\bar{R}_{\chic{Z}}(q^2) &=& 
\frac{\alpha_w}{c_w^2}
\bigg[q^2 - M_\chic{Z}^2 +\Re e\,\{\widehat{\Sigma}_{\chic{Z}\chic{Z}}(q^2)\}
\bigg]^{-1}
\nonumber\\  
\bar{s}_w^{2}(q^2) &=&  s_w^{2}\Biggl(1 - \frac{c_w}{s_w}\, 
\Re e\,\{\widehat{\Pi}_{\chic{A}\chic{Z}}(q^2)\}\Biggr) \,,
\label{RW}
\eea
where $\alpha_w = g_w^2/4\pi$, 
and $ \Re e\,\{...\}$ denotes the real part. 

In addition to being renormalization-group invariant, 
both quantities defined in  Eq.(\ref{RW})
are process-independent; 
$\bar{R}_{\chic{Z}}(q^2)$ corresponds to the $Z$-boson 
effective charge, while $\bar{s}_w^{2}(q^2)$  
corresponds to an effective mixing angle. 
We emphasize that 
the  renormalized  ${\widehat{\Pi}}_{ \chic{A}  {\chic  Z}} (0)$
{\it cannot}  form part of the NCR,  because 
it fails  to form a
renormalization-group invariant
quantity on its own.  
Instead,
${\widehat{\Pi}}_{\chic{A} {\chic Z}}  (0)$ must be combined with the
appropriate tree-level 
contribution (which  evidently  does not  enter into  the
definition of the NCR, since it is $Z$-mediated) 
in order to form the effective  
$\bar{s}_w^{2}(q^2)$ 
acting on the electron vertex,
whereas the finite 
NCR will be determined from the proper neutrino vertex only. 

After casting $\sigma^{(+)}_{\nu f}$ 
in terms of 
renormalization-group invariant blocks, one may 
fix $\nu = \nu_{\mu}$, 
and then consider three different choices for $f$: (i) 
right-handed electrons, $e_{\chic R}$; 
(ii) left-handed electrons, $e_{\chic L}$, and (iii) neutrinos, 
$\nu_{i}$ 
other than  $\nu_{\mu}$, i.e. $i=e,\tau$. 
Thus, we arrive at the system 
\bea
\sigma^{(+)}_{\nu_{\mu} \,\nu_i} &=& s \pi \bar{R}^2(0)
\nonumber\\
\sigma^{(+)}_{\nu_{\mu} \,e_{\chic R}}  &=& 
s \pi \bar{R}^2(0)\, \bar{s}_w^{4}(0) 
- 2 \lambda s_w^{2} \, 
\big< r^2_{\nu_{\mu}}\, \big>
\nonumber\\
\sigma^{(+)}_{\nu_{\mu} \,e_{\chic L}} &=& 
s \pi \bar{R}^2(0) \,
\bigg(\frac{1}{2} - \bar{s}_w^{2}(0)\bigg)^{2} \nonumber\\  
&& + \lambda (1-2 s_w^{2}) \, 
\big< r^2_{\nu_{\mu}}\, \big> 
\label{syst1}
\eea
where $\lambda \equiv (2\sqrt{2}/3) s \alpha \,G_{\chic F}$, 
$\alpha = e^2/4\pi$.
$\bar{R}^2(0)$, $\bar{s}_w^{2}(0)$, and  
$\big< r^2_{\nu_{\mu}}\, \big>$ are treated as three unknown 
quantities, to be determined from the above equations. 

To extract the experimental values of the quantities 
$\bar{R}^2(0)$, $\bar{s}_w^{2}(0)$, and $\big< r^2_{\nu_{\mu}}\, \big>$,
one must substitute  
in the above equations the experimentally
measured values for the differential cross-sections 
$\sigma^{(+)}_{\nu_{\mu} \,e_{\chic R}}$, 
$\sigma^{(+)}_{\nu_{\mu} \,e_{\chic L}}$,
and $\sigma^{(+)}_{\nu_{\mu} \,\nu_i}$. 
Thus, one would 
have to carry out three different pairs of experiments. 

{\it Acknowledgments}:
This work has been supported by the Spanish MCyT under the grant 
FPA2002-00612.

\end{document}